\documentclass[aps,preprint,amssymb]{revtex4}

\usepackage{epsfig}
\usepackage{bm}

\begin{document}
\renewcommand{\theequation}{\arabic{section}.\arabic{equation}}

%%%%%%%%%%%%%%%
\newcommand{\Div}{\textrm{div}}
\newcommand{\Rot}{\textrm{rot}}

\newcommand{\vect}{\vec}
\newcommand{\kvec}{\vect{k}}
\newcommand{\rvec}{\vect{r}}
\newcommand{\gvec}{\vect{G}}
\newcommand{\indx}{{n,\kvec}}
\newcommand{\rhat}{\hat{r}}

\newcommand{\beq}{\begin{eqnarray}}
\newcommand{\eeq}{\end{eqnarray}}

\newcommand{\half}{\frac{1}{2}}

\newcommand{\lbr}{\left( }
\newcommand{\rbr}{\right) }
\newcommand{\lbrla}{\left\langle }
\newcommand{\rbrra}{\right\rangle }

\newcommand{\lbrc}{\left\{ }
\newcommand{\rbrc}{\right\} }
\newcommand{\lbrs}{\left[ }
\newcommand{\rbrs}{\right] }
\newcommand{\lbrb}{\left. }
\newcommand{\rbrb}{\right. }
\newcommand{\lbrd}{\left. }
\newcommand{\rbrd}{\right. }
\newcommand{\lbrp}{\left| }
\newcommand{\rbrp}{\right| }

\newcommand{\contint}{\oint_{\gamma}}
\newcommand{\omint}{\oint_{\gamma}\frac{d\omega}{2\pi}}

\newcommand{\ylm}{Y_l^m}
\newcommand{\ylmp}{Y_{l'}^{m'}}
\newcommand{\yLM}{Y_L^M}
\newcommand{\ylms}{Y_l^{m*}}
\newcommand{\ylmps}{Y_{l'}^{m'*}}
\newcommand{\yLMs}{Y_L^{M*}}

\newcommand{\D}[2]{\frac{d #1}{d #2}}
\newcommand{\Dp}[2]{\frac{\partial #1}{\partial #2}}
\newcommand{\DF}[2]{\frac{\delta #1}{\delta #2}}

\newcommand{\DD}[2]{\frac{d^2 #1}{{d #2}^2}}
\newcommand{\DDpp}[3]{\frac{\partial^2 #1}{\partial #2 \partial #3}}
\newcommand{\DDp}[2]{\frac{\partial^2 #1}{{\partial #2}^2}}
\newcommand{\DDF}[3]{\frac{\delta^2 #1}{\delta #2 \delta #3}}

\newcommand{\sgn}{\mbox{sgn}}

\newcommand{\Pablo}{P. Garc\'{i}a-Gonz\'{a}lez }

\newcommand{\twf}[1]{{\tt #1}}
\newcommand{\itf}[1]{{\em #1}}
\newcommand{\bof}[1]{{\bf #1}}

\newcommand{\real}{\mbox{Re}}
\newcommand{\imag}{\mbox{Im}}

\newcommand{\aeq}{\simeq}

\newcommand{\totref}[1]{(\ref{#1}), page \pageref{#1}}

\newcommand{\bra}[1]{\lbrla #1 \rbrp}
\newcommand{\ket}[1]{\lbrp #1 \rbrra}
\newcommand{\braket}[2]{\lbrla #1 | #2 \rbrra}

\newcommand{\ignore}[1]{}

\newcommand{\allowbrk}{\allowbreak}
%%%%%%%%%%%%%%%%%%

\title{Wavepacket basis for time-dependent processes and its application to relaxation in
resonant electronic transport}

\author{Peter Bokes $^{a,b}$}
\email[]{E-mail: peter.bokes@stuba.sk}
%\thanks{}

\affiliation{$^a$ Department of Physics, Faculty of Electrical Engineering and Information
	     Technology, Slovak University of Technology, 812 19 Bratislava, Slovak Republic, \\
	     $^b$ European Theoretical Spectroscopical Facility (ETSF, {\tt www.etsf.eu})}
\date{\today}

\begin{abstract}
\noindent  
Stroboscopic wavepacket basis sets [P. Bokes, F. Corsetti, R. W. Godby, Phys. Rev. Lett. 101, 046402 (2008)]
are specifically tailored for a description of time-dependent processes 
in extended systems like non-periodic geometries of various contacts consisting 
of solids and molecules. The explanation of the construction of such a basis 
for two simple finite systems is followed  by a review of the general theory for 
extended systems with continuous spectrum. The latter is further elaborated with 
the introduction of the interaction representation which takes the full advantage of the 
time-dynamics built into the basis. The formalism is applied to a semi-analytical example 
of electronic transport through resonant tunnelling barrier in 1D. Through the 
time-dependent generalisation of the Landauer formula given in terms of the Fourier expansion 
of the transmission amplitude we analyze the temporal character of the onset 
of the steady-state. Various time-scales in this process are shown to be directly related to 
the energetic structure of the resonant barrier.

\end{abstract}

\maketitle

\section{Introduction}
\label{intro}

Time-dependent density functional theory (TDDFT) allows the formulation of first-principles simulations of 
many-electron problems in the language of non-interacting particles moving in the effective 
time-dependent Kohn-Sham potential. Actual solution of such a problem requires a particular representation,
or a basis set, for one-electron wavefunctions. There are many possibilities to choose 
from: system independent basis sets like the real-space representation~\cite{Castro06} 
and plane-waves~\cite{Qian06,Zhou09}, or more-or-less system-specific basis sets like various types 
of localized orbitals, either of Gaussian type~\cite{Schmidt93} or numerical 
orbitals~\cite{Brandbyge02,Rocha06}, ground state canonical Kohn-Sham orbitals corresponding 
to the studied system, maximally localized Wanier functions~\cite{Calzolari04} and possibly many other. 
The system independent basis set have the advantage of being 
general and not preferential to any particular system, which is balanced 
by the problem that typically one needs very large number of basis functions to obtain
converged results. This has been the main motivation in the development of basis sets
that are less general, more specialized for the description of the involved physics 
or chemistry at sufficient precision. With the latter the almost ubiquitous criterion
was their description of the ground states properties of the studied system.

In this paper we discuss a new type of basis sets that are specifically tuned 
for a description of time-dependent processes in extended systems like solids or non-periodic 
geometries of various contacts consisting of solids and molecules~\cite{Bokes08}. 
The major complications for these systems is their size. In the ground state one 
may hope to use ``nearsightedness'' of the electrons~\cite{Kohn96} and study the relevant 
questions using cluster or periodic models. In non-stationary processes this is not possible 
in general since electrons may traverse long distances and `see' arbitrarily remote 
parts of the system. A problem related to this, and one that is particularly relevant 
for the description of quantum transport within the  TDDFT is the possibility that the exact, 
or at least sufficiently precise exchange-correlation potential might need to 
be very nonlocal~\cite{Vignale09} even though the numerical importance of this effect in not clear 
at present~\cite{Sai05,Jung07}. 

One possible way is to study very large systems and look only at times
smaller than the time it takes for the electrons to realise that their box is in fact
finite~\cite{DiVentra04b,Bushong05,Qian06}.
Another possibility how to treat infinite systems is to employ embedding 
self-energies~\cite{Stefanucci04,Kurth05} that play the role of semi-infinite parts of the total system, 
often called electrodes, assumed to behave in a known way, and the rest, typically a small 
finite subsystem that can be treated fully within the TDDFT calculation. 
The stroboscopic wavepacket basis set gives a framework that 
avoids the use of embedding self-energies in principle and still treat the system as infinite. 
Individual basis functions are tailored to the time-dependence in the electrodes and 
the time-propagation of each electron there is analytical. Only in the regions where the system 
differs from the electrodes it is necessary to propagate the occupied orbitals numerically. 
Still, even here the wavepacket basis gives a very appealing picture of electrons being 
redistributed in between different wavepackets due to local disturbances in the effective potential.

In the following section we introduce the concept of stroboscopic construction of a basis set 
on a very simple examples: two level system consisting of $1s$ and $2s$ orbitals of Li atom 
and a subset of the eigenstates of Harmonic oscillator. In Sec.~\ref{infinite_sys} 
we review the formalism of the wavepacket basis for any Hamiltonian with a continuum 
spectrum which is in Sec.~\ref{td-basis} further elaborated into interaction representation 
based on the wavepacket basis. The following two sections give an example of the use of the basis
on transport of electrons through a simple resonant potential barrier. We find a simple
and direct relation between the Fourier analysis of a transmission amplitude and the 
temporal behaviour of the relaxation into the current-carrying steady state.

\section{Stroboscopic wavepacket basis}
\label{strobo_basis}
\subsection{Finite systems}
\label{finite_sys}
Let us consider a two level system, e.g. the canonical 1s and 2s orbitals obtained within the ground
state DFT calculation of the Lithium atom, $\ket{1s}$ and $\ket{2s}$ with eigenenergies $e_1$ and $e_2$. 
To obtain the ground state density, we occupy the first orbital twice and the second one once 
with the result
\begin{equation} 
	n(\vec r) = 2 | \braket{\vec r}{1s} |^2 + | \braket{\vec r}{2s} |^2.
\end{equation}
In principle, any unitary rotation between these occupied canonical orbitals gives identical density.
We will keep one electron, say with its spin down, in the 1s orbital, but the other two, both with 
spin up, will be placed into new orbitals,
\begin{eqnarray} 
	\ket{g} &=& \frac{1}{\sqrt{2}} ( \ket{1s} + \ket{2s} ) \\
	\ket{u'} &=& \frac{1}{\sqrt{2}} ( \ket{1s} - \ket{2s} )
\end{eqnarray}
This is, of course, a well known fact. New and interesting point of view arises
if we view the state $\ket{u'}$ as the result of time-propagation of the 
$\ket{g}$\footnote{Throughout the paper we use the atomic units a.u., where $m=\hbar=e=1$.}
%When using footnotes, please follow the command \verb|\footnote[X]{The footnote text}| where 
%for X the numbers from 1 to 9 should be input in succession.\\
\begin{equation} 
	e^{-i\hat H \tau} \ket{g} = \frac{1}{\sqrt{2}} e^{-ie_1\tau}  \left( 
		\ket{1s} + e^{-i(e_2 - e_1)\tau} \ket{2s} \right) = e^{-ie_1\tau} \ket{u'} 
	\textrm{ for } \tau = \frac{\pi}{e_2-e_1}.
\end{equation}
This means, we could consider orbitals $\ket{g}$ and $\ket{u} = e^{-i\hat{H} \tau} \ket{g}$
as two orthogonal states which can be filled with electrons. This rather trivial example 
clearly demonstrates the principal idea of the stroboscopic basis set: to obtain orthogonal 
basis by means of a propagation from suitably chosen initial state.

The second example where such a construction works for discreet spectra is harmonic oscillator.
Consider $N$ consecutive states with eigenenergies $e_n=\hbar \omega_0 n, \quad n=n_0,\dots,n_0 + N-1$
and eigenstates $\ket{\phi_n}$. 
We first choose the {\it initial state} as the equal combination of the eigenstates, 
\begin{equation}
	\ket{\psi_0} = \frac{1}{\sqrt{N}} \sum_{n=n_0}^{n_0+N-1} \ket{\phi_n}. \label{eq-finite-10}
\end{equation}
In principle, the phases of each of the coefficient in the initial state are arbitrary, which 
reflects certain freedom in choosing a particular basis.

The remaining members of the new basis are obtained through propagating the initial state with time step 
$\tau = 2\pi / (N \omega_0)$,
\begin{equation}
	\ket{\psi_m} = e^{-i\hat{H} m \tau} \ket{\psi_0}, \quad m=0,1,...,N-1.  \label{eq-finite-11}
\end{equation}
It is a simple excercise to show that this set is orthogonal, $\braket{\psi_m}{\psi_m'} = \delta_{mm'}$,
and spans identical subspace as the chosen eigenstates. The new basis carries certain resemblance 
with the coherent states used extensively within the quantum optics~\cite{Glauber63}. 
Namely, both lead to almost classical interpretation of the dynamics 
of a single quantum particle, e.g. the contribution of a single electron to the total density is 
oscillating from right to the left of the parabolic potential with the period $T=2\pi/\omega_0$.
% \begin{equation} 
% 	n_1(x) = |\braket{\psi_1}{\psi_1}|^2 = 
% \end{equation}
On the other hand, in contrast with the coherent states, the time-propagated states 
by construction consists only from eigenstates with energies within given interval
which allows for exact description of non-interacting many-electron ground states 
or simple non-equilibrium models, as it is described in the following sections.

%When using footnotes, please follow the command \verb|\footnote[X]{The footnote text}| where 
%for X the numbers from 1 to 9 should be input in succession.\\

\subsection{Infinite systems}
\label{infinite_sys}
Systems of infinite extent in space have not only discreet but most importantly 
also continuous spectrum of eigenvalues. In our previous work~\cite{Bokes08} we have shown that 
the states with eigenvalues from chosen interval of energies $(\epsilon,\epsilon+\Delta \epsilon)$ 
in the continuous spectrum can be unitarily transformed into wavepacket basis, in a similar 
way to the examples in the previous section. In the following we will review the basic 
steps of this formalism.

Let the system of noninteracting particles under consideration be initially 
characterised by a {\it reference} Hamiltonian $\hat{H}$ 
with a continuous part of its spectrum covered with intervals of energies 
$(\epsilon_n,\epsilon_{n+1}), \quad n=0,1,...$ to which we refer here as the {\it energy bands}. 
We will assume that each energy within given band is $N_n$ times degenerate. 
The eigenstates of the reference Hamiltonian,
$$
        \hat{H} \ket{\epsilon,\alpha} = \epsilon \ket{\epsilon,\alpha},
$$
need to be taken normalized to the delta-function of their energies,  
\begin{equation}
        \braket{\epsilon',\alpha'}{\epsilon,\alpha} =  \label{eq-f1}
                        \delta(\epsilon-\epsilon') \delta_{\alpha,\alpha'}.
\end{equation}

From the above set of eigenstates we can generate the wave-packet basis set (WPB) by first choosing 
the {\it initial set} of wave-packets (WPs)
\begin{equation}
        \ket{n,0,\alpha} =  \label{eq-f2}
        \frac{1}{\sqrt{\Delta \epsilon_n}}
        \int_{\epsilon_n}^{\epsilon_{n+1}} d \epsilon'
        \ket{\epsilon',\alpha}, \quad n=0,1,2,...
\end{equation}
for each energy band $\left\{ (\epsilon_n,\epsilon_{n+1}) \right\}_{n}$ and each 
value of the degeneracy index $\alpha$. $\Delta \epsilon_n=\epsilon_{n+1}-\epsilon_{n}$
represents the width of all the considered energy bands.

The construction of the WPB is completed by forward and backward time propagation 
of each member of the initial set
\begin{equation}
        \ket{n, m,\alpha} =  \label{eq-f3}
        e^{-i\hat{H} m \tau_n} \ket{n,0,\alpha}, \quad
        m = \pm 1, \pm 2, \ldots
\end{equation}
by regular, band-dependent time steps $\tau_n=2\pi / \Delta \epsilon_n$. This choice of time steps
guarantees orthonormality of consecutive wave-packets within each band since,
\begin{eqnarray}
        \braket{n,m,\alpha}{n,m',\alpha} 
		&=&  \frac{1}{\Delta \epsilon_n} \int d \epsilon d \epsilon' \bra{\epsilon,\alpha} 
			e^{-i(\epsilon' m' - \epsilon m ) \tau } \ket{\epsilon', \alpha} \nonumber \\
		&=&  \frac{1}{\Delta \epsilon_n} \int d \epsilon e^{-i\epsilon (m' - m ) \tau } 
			= \delta_{m,m'},
\end{eqnarray}
where we have used the delta-function normalisation (\ref{eq-f1}). The same equation also 
guarantees that WPs with different degeneracy index $\alpha$, as well as WPs from
disjunct energy bands are mutually orthogonal. Collecting together all such WPs together with 
the discreet eigenstates (if they are present) of the reference Hamiltonian 
we form a complete orthogonal set of states.
The regular time steps between individual WPs remind a view through a stroboscope 
on a continually moving single wavepacket. This analogy led us to refer to such a complete orthogonal 
set as the {\it stroboscopic wavepacket basis}. 

A particular example of the stroboscopic basis are the orthogonal wavepackets 
constructured from the plane waves and used first by Th. Martin and R. Landauer~\cite{Martin92}.
In fact, this work served as a motivation for generalisation into the stroboscopic 
construction in the present form.

There are several very attractive properties of this basis that we would like 
to emphasize: (1) instead of continuous and delta-function normalised states we can work 
with countable even though infinitely many states that are localized in space and normalised 
to one; (2) occupying all states from all energy bands below the Fermi energy, $E_F$, we recover 
identical many-particle ground state of non-interacting particles characterised by 
the reference Hamiltonian; (3) while the total density of such many-particle ground state 
is time-independent as it should, electron in each WP state in the energy interval 
$(\epsilon_n,\epsilon_n+\Delta \epsilon_n)$ can be interpreted as making transition from $m$-th 
WP to $m+1$-th WP in time $\tau_n$. As it will be demonstrated in Section \ref{resonant_tunel}, 
all these properties can be very useful for quick and insightful description of non-equilibrium 
processes in many-electron systems.

\subsection{Time-dependent wavepacket basis}
\label{td-basis}
In the previous section the wavepacket basis set was constructed in terms of stroboscopic 
snapshots of time-propagation of initial set of suitably chosen wavepackets. Clearly, 
this construction is valid for {\it any} initial time. This can be exploited even further  
within the interaction representation where each time-dependent basis function
is an evolving wavepacket, and different orthogonal basis functions are time-shifted with 
respect to each other by $\tau$. This means that any wavefunction $\phi(x,t)=\braket{x}{\phi,t}$
occupied by an electron can be expanded as 
\begin{eqnarray}
        \braket{x}{\phi,t} = \sum_{i,n,\alpha}  \braket{x}{i,\alpha,n;t} c_{i,\alpha,n}(t)
\end{eqnarray}
where $i$ runs over different energy bands, $\alpha$ through different degenerate states 
within the band and $n$ over different orthogonal wavepackets from the given
band $i$.  Due to their very definition through time-evolution these time-dependent basis function 
have the property 
\begin{equation}
        \braket{x}{i,\alpha,n;t+\tau_i}  = \braket{x}{i,\alpha,n+1;t},
\end{equation}
for any real time $t$, $\tau_i$ given by the width of the energy band $i$ and any integer $n$
indexing the various WPs within the band.

There is not much gained from using this particular representation if the actual Hamiltonian 
of the system is the reference Hamiltonian which generates the wavepacket basis
at all times. In such a case the coefficients $c_{i,\alpha,n}(t)$ are either equal 1, if 
the considered band $i$ is below the Fermi energy, or equal 0, if the band $i$ is above the Fermi energy. 
However, when the actual Hamiltonian locally differs from the reference Hamiltonian  
at times $t>t_0$, i.e. the Hamiltonian is 
\begin{equation}
	\hat{H} = \hat{H}_0 + \hat{V}(t)
\end{equation}
where $\hat{V}(t) \neq 0 $ only in finite part of the overall system, $\Omega \subset \bm R^3$, and 
$t>t_0$, the wavefunctions' coefficients will evolve according the Schroedinger equation 
in the interaction picture
\begin{eqnarray}
        i \Dp{~}{t} c_{j,\alpha,m}(t) &=& \sum_{i,\alpha',n} \bra{j,\alpha,m;t} \hat{V}(t) 
		\ket{i,\alpha',n;t} c_{i,\alpha',n}(t). \label{eq-int-Sch}
\end{eqnarray}
Since the perturbation $\hat{V}(t)$ is assumed to be nonzero only in $\Omega$, and the 
wavepackets are localised in space, the above matrix elements are nonzero only for those 
WPs that have significant amplitude in $\Omega$. The coefficient  $c_{i,\alpha,n}(t)$ 
for times before the corresponding basis function $\braket{x}{i,\alpha,n;t}$ is evolved 
by the reference Hamiltonian into the region $\Omega$ will be equal 1, reflecting 
the ground state for $t<t_0$ imposed by the reference Hamiltonian in the past. During 
the following time, when the matrix elements of this WP basis function are nonzero, 
the coefficient $c_{i,\alpha,n}(t)$ will change according to Eq.~(\ref{eq-int-Sch}). Finally, 
once the WP basis function $\braket{x}{i,\alpha,n;t}$ departs from $\Omega$, the coefficient 
will freeze and will be kept constant once again.

It is clear from this discussion that the use of the interaction picture 
representation of the stroboscopic wavepacket basis allows us to translate 
the time-dependent quantum-mechanical problem of a system with infinite extent into one 
that demands numerical solution of finite number of dynamical equations (\ref{eq-int-Sch}). 
We will demonstrate this strategy in the following sections where we address the behaviour 
of the relaxation into a current carrying steady-state for a model resonant nano-junction 
at moderate-to-long time-scales.

\section{Electronic transport through resonant tunnelling barrier}
\label{resonant_tunel}

\subsection{Definition of the model}
\label{model}

\begin{figure}[ht]
  \begin{center}
   \includegraphics{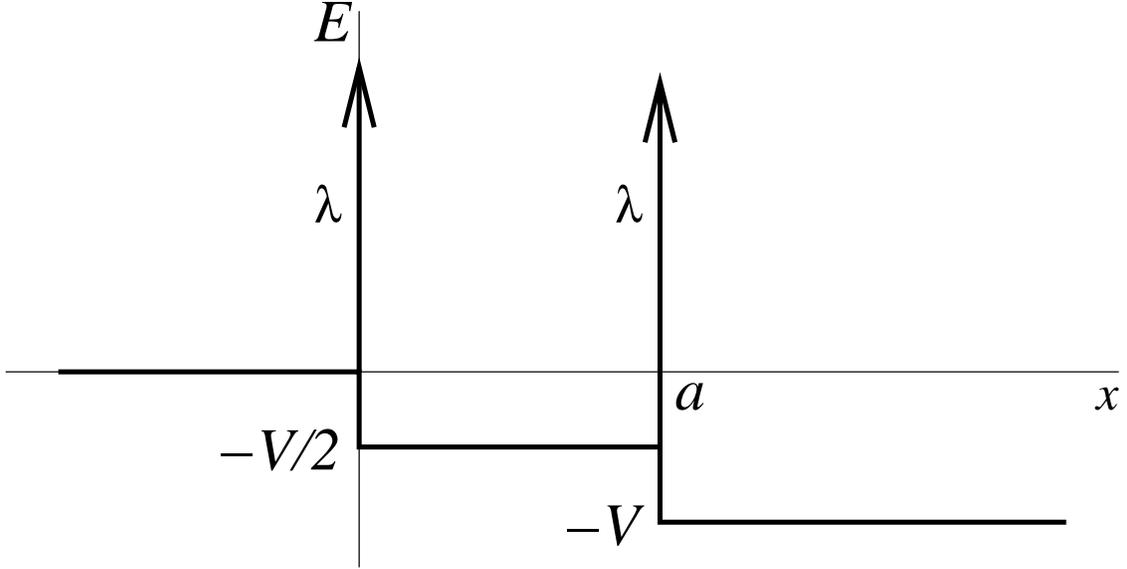}
   \caption{Form of the model resonant potential at nonzero bias $V$. The value of the 
	potential inside the barrier is chosen to be halfway between the bias.} \label{fig-1}
  \end{center}
\end{figure}

For the purposes of demonstration of the use of the time-dependent stroboscopic basis
we will study a very simple model of resonant tunnelling. In many respects, such a model
can be viewed as a prototype of many interesting phenomena observed in the nanoscopic 
transport through molecules attached to metallic electrodes~\cite{DiVentra00,Nitzan03,Koentopp05}. 
The time-dependence in tunnelling has been addressed using various wavepackets by many authors in 
the past, e.g.~\cite{Buttiker82, Guo88, Jauho89, Stovneng91, Landauer94, Garcia-Calderon01}.
The distinguishing contribution of the present treatment is the correct many-electron occupation 
corresponding to finite bias and low temperatures and hence its direct consequences for the 
current-voltage characteristics.
The system will be strictly one dimensional, with two delta-function potential barriers 
localised at $x=0$ and $x=a$ with equal strengths given by a parameter $\lambda$ (see Fig.~\ref{fig-1}). 

Applying voltage to such a model results in decreasing the constant potential to the 
right of the double barrier by a value $-V$. The value of the potential 
in between the two potential barriers is chosen to be $-V/2$. The physics of this choice is 
that the electrons inside the double barrier will effectively screen the field and the 
total drop in potential will be symmetrically split into the two barriers where the electronic
density is low and hence the screening is weaker.

The simplicity of this model allows for analytical construction of its right- and left-
going scattering eigenstates for any positive energy $\epsilon$, e.g. for the right-going state we have
\begin{equation}
	\phi_k^R(x) = \frac{1}{\sqrt{2\pi k}} \left\{ \begin{array}{cc}  \label{eq-phi_R}
			e^{ikx} + r(e) e^{-ikx}, k=\sqrt{2\epsilon} &  x < 0 \\
			C_1(\epsilon) e^{iqx} + C_2(\epsilon) e^{-iqx}, q=\sqrt{2(\epsilon+V/2)} & 0<x<a \\
			t(e) e^{i\kappa x},  \kappa=\sqrt{2(\epsilon+V)} & x>a
			\end{array} \right. , 
\end{equation} 
where all the functions $r(\epsilon), C_1(\epsilon), C_2(\epsilon)$ and $t(\epsilon)$ can 
be obtained from the conditions on continuity the wavefunction and finite discontinuity of 
its derivative at $x=0$ and $a$, and are simple functions of the energy $\epsilon$ of 
the considered  state. Only the transmission amplitude, $t(\epsilon)$, is directly relevant 
for the present work and its functional form is 
\begin{equation} 
	t(\epsilon) = \frac{4kq}{
        (\kappa + q + 2i\lambda)(k+ q + 2i\lambda) e^{i(\kappa - q)a}
         - (\kappa - q + 2i\lambda)(k-q+2i\lambda) e^{i(\kappa + q)a}}.
\end{equation}
The prefactor of the scattering state $\phi_k^R(x)$, $1/\sqrt{2\pi k}$, guarantees correct 
normalisation given by Eq.~(\ref{eq-f1}). The right- and left- going scattering states at the same energy
are examples of two degenerate continuum eigenstates. In the language of the general formulation 
of the wavepacket basis~[Eqs.~(\ref{eq-f2}),~(\ref{eq-f3}) ], their indices $R$ and $L$ represent 
two possibilities 
of the degeneracy index $\alpha$. The resulting WPs for $\alpha=R$ and $L$ will be arriving 
to the barrier from the left or right respectively.

\begin{figure}[ht]
  \begin{center}
   \includegraphics{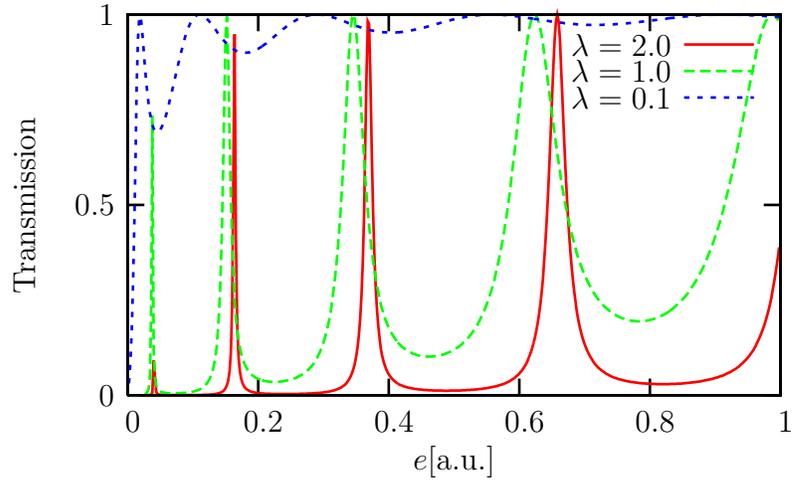}
   \caption{Transmission function for the model resonant potential at zero bias. $E_F=1.0$
	guarantees that several resonances are occupied in equilibrium, before the bias is 
	switched on.} \label{fig-2}
  \end{center}
\end{figure}

The particular values of the parameters will be chosen in the following way:
the Fermi energy is set to $E_F=1.0$, the strength of the barriers will be varied through 
values $\lambda=0.1,1.0$ and $2.0$. 
We will use two different distances between the barriers, $a_1=9.5$ and $a_2=10.5$ to 
clearly demonstrate the different types of relaxation behaviour. Both of these 
are above the Fermi wavelength $\lambda_F=2\pi/(\sqrt{2 E_F}) = 4.44$ to allow 
for existence of several resonances within the energy interval $(0,E_F)$ at zero bias.
In Fig.~\ref{fig-2} we show the transmission of such a potential for nonzero 
bias $V$ and several values of $\lambda$. 

\subsection{Non-linear response to abrupt switching-on of the voltage}
\label{relaxation}

In equilibrium, all from the right- and left- moving wavepackets 
arising from the energy bands below the Fermi energy will be occupied so that overall current is zero. 
This is not the case in non-equilibrium situation, and nonzero current can be obtained.
The calculation of the current alone can be done using the free electron wavepackets (FWP), 
\begin{equation}
\bra{x}\left. i,\alpha,n; t \right) = \int_{\epsilon_n}^{\epsilon_{n+1}} 
\frac{d \epsilon}{\sqrt{2\pi k}} e^{i\alpha k x} e^{-i\epsilon t}, \quad k = \sqrt{2\epsilon}
\end{equation}
i.e. the wavepackets where the reference Hamiltonian is that of free electrons 
$\hat{H}_0 = - (1/2) d^2/dx^2$. This Hamiltonian characterises the regions to the right and left 
of the double barrier. (We will use the round bracket to distinguish the FWP from a 
general WP $\ket{i,\alpha,n; t}$.) The degeneracy index $\alpha$ for FWPs is equal to $+$ or $-$ 
and indicates the right- and left- moving FWPs respectively. Within this representation 
the current due to the energy band or the {\it bias window}  $(E_F-V,E_F)$ 
(the corresponding time step between neighbouring FWPs is $\tau_V=2\pi/V$) is just a simple counting: 
number of electrons that occupy $n$-th right-going FWP divided by the time it takes for 
an electron to move from $n$-th to the $n+1$-th FWP,
\begin{equation} 
	I_n(t) = \frac{N_n(t)}{\tau_V} = \frac{N_n(t)}{2\pi} V,
\end{equation} 
where $N_n(t)$ is the average occupation of the $n$-th FWP.

Immediately after the switching-on the bias, the evaluation of the current at a chosen position 
of the FWP and time needs to be analyzed numerically since the time-dependent 
Hamiltonian will occupy various unoccupied bands of higher energy and these would 
also contribute to the current. However, it is possible to arrive at a simple expression 
for the occupations $N_n(t)$ for moderate to long time-scales when only WPs 
in the bias window $(E_F-V,E_F)$ contribute to the current. Under these conditions 
the average occupation is given in terms of the overlaps between the scattering wavepackets 
corresponding to the {\it final} Hamiltonian which includes the barrier and the bias potential,
and the FWPs~\cite{Bokes08},
\begin{equation}
	N_n(t) = 2\sum_{m=0}^{-\infty} |\left(n,+,i \right. \ket{i,R,m;t} |^2, \label{eq-Nnt}
\end{equation}
where 
\begin{equation} 
	\left(i,+,n \right. \ket{i,R,m;t}  = \frac{1}{V} \label{eq-transmission-time}
	\int_{E_F-V}^{E_F} d \epsilon t(\epsilon) \exp\{- i \epsilon (t + (m-n)\tau_V)\}
\end{equation}
The $i$-th energy band is understood to be the bias window, the factor $2$ stands for spin 
degeneracy. The Eq.~(\ref{eq-transmission-time}) is strictly valid only for intermediate 
and long times times, $t> 1/E_F$, as well as only for FWPs that are sufficiently distant 
from the resonant barriers, $n>a/(v_F \tau_V)$. The simplicity of Eqs.~(\ref{eq-Nnt})
and (\ref{eq-transmission-time}) is striking.  The Eq.~(\ref{eq-transmission-time}) is simply 
a Fourier analysis of the transmission on the interval of energies that fall within the bias window. 
Let us introduce the coefficients of this Fourier series as 
\begin{equation} 
	t_\eta(l) = \int \frac{d \epsilon}{V} t(\epsilon) e^{-i\epsilon \eta}
                \exp\{-i \epsilon (l\tau_V)\} \label{eq-Four-coef}, 
\end{equation}
where the time is split into the integer multiple of the time step $\tau_V$ and the rest, 
$t = N_t \tau_V + \eta, \quad 0<\eta<\tau_V$. Using this convention we can write 
the total current at the position of $n$-th FWP as 
\begin{equation} 
	I_n(t) = \sum_{m=0}^{-\infty} |t_\eta(N_t+m-n)|^2,
\end{equation} 
which can be readily evaluated. It turns out that this sum converges rapidly:
for the here studied resonant barrier the results for the times show in the 
Fig. \ref{fig-3} and \ref{fig-4} are converged if we take just 15-20 terms.
This is an indication that the number of needed WPs to achieve convergence of results
does not need to be large. Similar observation has been found in convergence 
of the ground states density for square potential barrier~\cite{Bokes-EPAPS}.

\begin{figure}[ht]
  \begin{center}
   \includegraphics{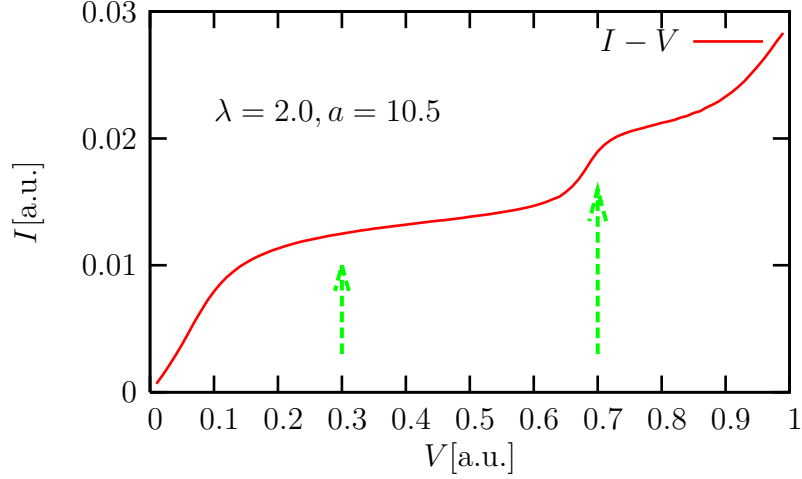}
   \caption{The steady-state current-voltage characteristics of the resonant barrier. Rapid
	increases in current correspond to an emergence of next resonance within 
	the bias window. The green arrows indicate voltages for which we have studied 
	the time-dependent approach to the here indicated values of current.} \label{fig-2b}
  \end{center}
\end{figure}

\begin{figure}[ht]
  \begin{center}
   \includegraphics{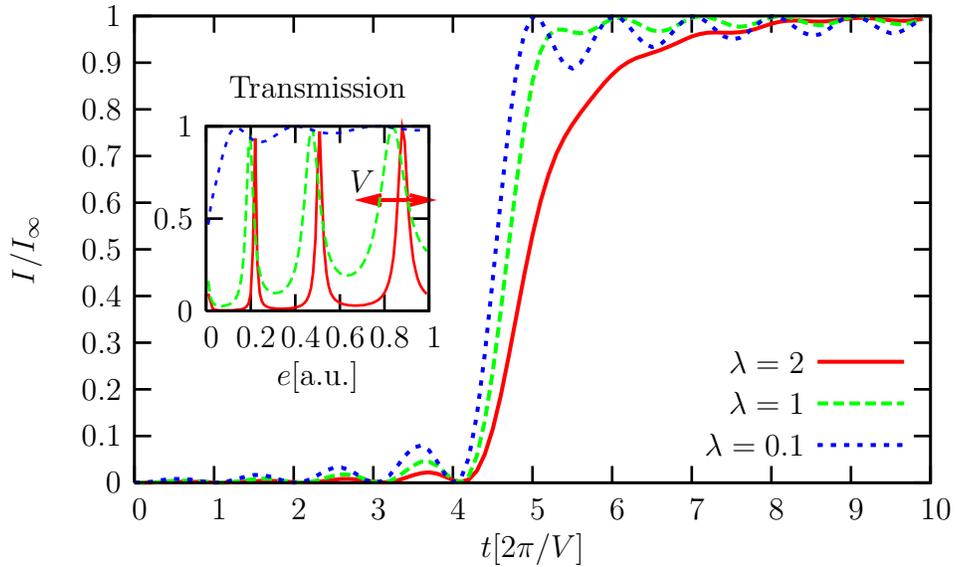}
   \caption{Relaxation of the current towards its steady-states value, $I_\infty$, for the case
	when single resonance is present within the bias window. Varying the strength 
	of the barriers alters the character for oscillatory to exponential. The inset 
	shows the transmission spectrum at this value of the bias with the resonance 
	present in the bias window.} \label{fig-3}
  \end{center}
\end{figure}

\begin{figure}[ht]
  \begin{center}
   \includegraphics{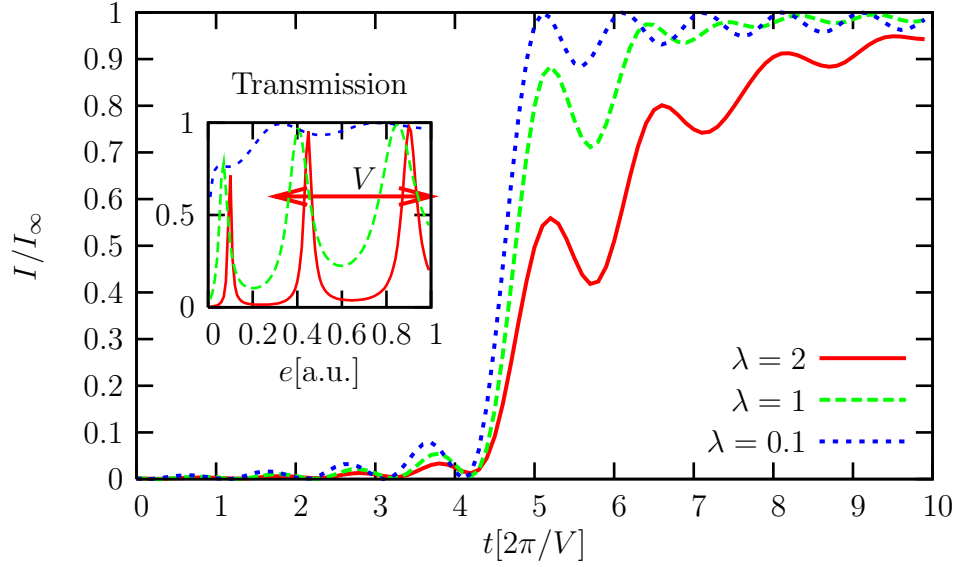}
   \caption{Relaxation of the current towards its steady-states value, $I_\infty$, for the case
	when two resonances are present within the bias window. Varying the strength 
	of the barriers alters the character for oscillatory to exponential with superimposed 
	oscillations that can be ascribed to the energy distance between the two resonances. 
	The inset shows the transmission spectrum at this value of the bias and the two resonances
	present in the bias window.} \label{fig-4}
  \end{center}
\end{figure}

We now proceed to discuss the results for two different values of applied voltage. 
First we choose such a voltage that only one resonance appears in the bias window 
$(E_F,E_F-V)$. Next we increase the voltage and shorten the distance between the two 
barriers to $a_2=9.5$ to capture two resonances. In the long time limit, the current 
goes into its steady-state value $I_\infty$ given by the Landauer formula~\cite{Bokes08}
\begin{equation} 
	I_\infty = \frac{1}{\pi} \int_{E_F-V}^{E_F} d \epsilon |t(\epsilon)|^2.
\end{equation}
The resulting $I-V$ characteristic for our model is given in the Fig.~\ref{fig-2b}. 
Whenever new resonance enters the bias window $(E_F-V,E_F)$ a rapid increase in the current appears, 
which is a well known feature in the transport through molecular nano-junctions~\cite{Nitzan03}.

The resulting dependence of the current on time, studied at the location of the $n=4$-th FWP
($n=0$ corresponds to FWP located behind the position $x=a$) for the bias voltage
$V=0.3$ is shown in Fig.~\ref{fig-3}. In this case only a single resonance enters 
the bias window, as it is indicated in the inset of the Fig.~\ref{fig-3}. It takes time 
$t=4\tau_V$ for the front with nonzero current to appear at the $4$-th FWP basis function 
in the right lead, and the steady-state current is set in with damped oscillations 
whose period is given by the time step $\tau_V$. This is particularly clear for 
transmissive enough barriers ($\lambda=0.1$). Such a behaviour is {\it not just an artefact} 
of the WP basis and the choice of the energy band of width $V$; identical voltage-dependent 
oscillations were found using non-equilibrium Green's function formalism~\cite{Stefanucci04}. 
Increasing the strength of the barriers, i.e. increasing the parameter lambda, the oscillations 
are suppressed and instead, new time-scale appears in the functional form of the time-dependence: 
a slow exponential approach towards the steady-state 
current. The time-constant of this approach is given by the width of the resonance 
present within the bias window $V$. This can be confirmed by inspection of 
the transmission spectrum (inset of the Fig.~\ref{fig-3}).

If two resonances are present within the bias window ($V=0.7$, Fig.~\ref{fig-4}), the behaviour 
of the current relaxation is similar to the previous case for highly transmissive barriers. 
However, increasing $\lambda$ and hence making the barriers more opaque brings 
in a new time-scale: the slow exponential approach to the steady-state value determined by the 
widths of the resonances is now superposed with oscillations with period slightly larger 
than $\tau_V$. This can be ascribed to the energy difference between the two resonances 
that are both within the bias window (the inset of the Fig.~\ref{fig-4}).

The example of electronic transport through resonant barrier shows how the time-dependent 
expression for the current in terms of the Fourier decomposition of the transmission, 
Eq.~(\ref{eq-Four-coef}) allows for a very simple analysis of the time-dependent processes 
involved in the relaxation of the current towards its steady-state value.

\section{Conclusions} 
\label{conclusions}
The stroboscopic wavepacket basis represent a novel basis set specifically tailored 
for a description of time-dependent processes in extended systems 
We have explained the idea of the stroboscopic construction on a simple two level 
system and a finite system consisting of subset of eigenstates of the harmonic oscillator. 
In the following we have reviewed the general theory 
for construction of the stroboscopic wavepacket basis for an extended system with continuous 
spectrum. We have introduced interaction representation within 
this basis which practically reduces the number of dynamical equations that need 
to be solved to finite number. 

We have applied the time-dependent formalism to semi-analytical example of electronic 
transport through resonant tunnelling barrier in 1D. The use of stroboscopic basis 
allows for physically appealing and at the same time mathematically exact formulation 
of the transport. Furthermore, through the time-dependent generalisation of the Landauer 
formula in terms of Fourier analysis of the transmission amplitude we could identify 
the temporal behaviour of the establishment of the steady state with nonzero current where
various time-scales could be put directly into relation with the energetic structure of the
resonant barrier. 

\acknowledgements
The author wishes to acknowledge fruitful discussions with R. W. Godby, J. T\'{o}bik and 
P. Die\v{s}ka. This research has been supported by the NANOQUANTA EU Network of Excellence 
(NMP4-CT-2004-500198) and the Slovak grant agency VEGA (project No. 1/0452/09).
.

%The references should start on their own page.
\clearpage

%\bibliography{References}
%\bibliographystyle{pccp-bibtex}

%\begin{thebibliography}{99}
%\bibitem{marker}
%Author(s), {\it Journal title}, Year, {\bf Volume}(Issue number), First page number.
% BibTeX users can use a style-file for PCCP, which can be found on the CTAN internet site 
% of LaTeX: http://www.tex.ac.uk/tex-archive/biblio/bibtex/contrib/chem-journal/pccp.bst
%\end{thebibliography}

%\clearpage
%The tables should be submitted normally after the reference list, starting on a separate page.
%\begin{table}
%\begin{center}
%\caption{\label{tab1}Insert table caption here}
%\begin{tabular}{llll}
%\hline
%Header 1 & Header 2 & \multicolumn{2}{l}{Header 3}\\ \cline{3-4}
%&&Subheader 1 & Subheader 2 \\ \hline
%Column 1 & Column 2 & Column 3 & Column 4\\
%Column 1 & Column 2 & Column 3 & Column 4\\
%\hline
%\end{tabular}
%\end{center}
%\end{table}

%Please compile a list of all figure captions on a separate page:
\clearpage
\begin{list}{}{\leftmargin 2cm \labelwidth 1.5cm \labelsep 0.5cm}

\item[\bf Fig. 1] Form of the model resonant potential at nonzero bias $V$. The value of the 
	potential inside the barrier is chosen to be halfway between the bias.

\item[\bf Fig. 2] Transmission function for the model resonant potential at zero bias. $E_F=1.0$
	guarantees that several resonances are occupied in equilibrium, before the bias is 
	switched on.
\item[\bf Fig. 3] The steady-state current-voltage characteristics of the resonant barrier. Rapid
	increases in current correspond to an emergence of next resonance within 
	the bias window. The green arrows indicate voltages for which we have studied 
	the time-dependent approach to the here indicated values of current.

\item[\bf Fig. 4] Relaxation of the current towards its steady-states value, $I_\infty$, for the case
	when single resonance is present within the bias window. Varying the strength 
	of the barriers alters the character for oscillatory to exponential. The inset 
	shows the transmission spectrum at this value of the bias with the resonance 
	present in the bias window.

\item[\bf Fig. 5] Relaxation of the current towards its steady-states value, $I_\infty$, for the case
	when two resonances are present within the bias window. Varying the strength 
	of the barriers alters the character for oscillatory to exponential with superimposed 
	oscillations that can be ascribed to the energy distance between the two resonances. 
	The inset shows the transmission spectrum at this value of the bias and the two resonances
	present in the bias window.

\end{list}

\clearpage

\end{document}